\documentclass[twocolumn,prl,showpacs]{revtex4}

\usepackage[german,english]{babel}
\usepackage{amssymb}
\usepackage{amsfonts}

\begin{document}

\title{The future is noisy: on the role of spatial fluctuations
in genetic switching}

\date{\today}

\author{Ralf Metzler}
\email{metz@physik.uni-ulm.de}
\affiliation{Department of Physics, Massachusetts
Institute of Technology, 77 Massachusetts Ave., Rm 12-109, Cambridge,
Massachusetts 02139, U.S.A.}

\begin{abstract}
A genetic switch may be realised by a certain operator sector on the DNA
strand from which either genetic code, to the left or to the right of this
operator sector, can be transcribed and the corresponding information
processed. This switch is controlled by messenger molecules, i.e., they
determine to which side the switch is flipped. Recently, it has
been realised that noise plays an elementary role in genetic switching,
and the effect of number fluctuations of the messenger molecules have been 
explored. Here we argue that the assumption of well-stirredness
taken in the previous models may not be sufficient to characterise the
influence of noise: {\em spatial} fluctuations play a non-negligible part in
cellular genetic switching processes.
\end{abstract}

\pacs{87.15.Ya,87.16.-b,05.40.-a,02.50.-r}

\maketitle

The Darwinistic theory of evolution through mutation and selection is
based on occasional errors in the process of biological reproduction
\cite{darwin}.
At the same time, a biological system has to be locally stable in the sense
that a mutation occurs only once within a large number of generations of a
species. Biological systems, faced with thermal activation at room
temperature therefore have to find a way to minimise the influence of noise.
Particular interest in this concern is focused on genetic switches, a
relatively simple unit in a biological cell at hand of which the effects
of noise can be studied.

Genetic switches constitute a part of the centre of command of biological
cells. By deciding which of two genetic codes within the corresponding DNA
section is transcribed, the switch prompts the production of certain molecules
inside the cell, and thus it controls the subsequent reactions and the
feedback: the genetic switch governs the future state
of the cell. In fact, the associated interplay of different chemical reactions
resembles a logical electric circuit, and biochemists have therefore coined
the notion of genetic circuitry \cite{ptashne,alberts,arkinross}.
Genetic switches, depending on their functional task, come either
non-cooperative or cooperative. Roughly speaking, cooperative switches will
always be concerned with the control of vital processes such as reproduction,
non-cooperativity is associated with less precision and concerns processes
like respiration. For a large number of cellular systems, the steric and
kinetic aspects of the biochemistry of the molecules involved in
genetic switching have been explored to great detail, and it has been
investigated how and where cooperativity originates.
In general, cooperative switches are much more robust
against noise \cite{ptashne,alberts}.

The paradigm model for a cooperative genetic switch is the system made up by
the host bacterium {\em Escherichia coli} which is infected by the parasitic
bacteriophage T4 ($\lambda$-phage). $\lambda$ injects
its own DNA into the host cell where the $\lambda$-DNA fuses with the bacterium
DNA. Thus, $\lambda$ is able to abuse the host cell facilities to either remain
dormant and get reproduced along with the bacterium ({\em lysogeny}), or fuse
new $\lambda$-phages by the help of the host cell's miniature chemical plants,
the process of {\em lysis}. The latter eventually leads to an array of a large
number of new $\lambda$ inside {\em E.coli}. Finally the host cell bursts and
releases a swarm of new $\lambda$'s \cite{ptashne,alberts}. Which of the two
paths, lysis or lysogeny, is followed, is determined by a genetic switch which,
in turn, is triggered by messenger molecules which we call repressor
R (the protagonist lysogeny agent) and aggressor $\overline{{\rm
R}}$, the antagonist lysis agent.

The typical, overall number of messenger molecules within a biological cell
like {\em E.coli} is relatively small, ranging from a few to some 100. On the
one hand, the chemical processes involved in the synthesis and degradation of
the messenger molecules are noisy and, to some approximation, described by a
master equation \cite{vankampen}. Noise means that at some given time there
is a surplus in the molecule synthesis in respect to the degradation, and the
overall number of the respective molecule increases; and vice versa. That
means that even in a state which is stationary on average, fluctuations in the
number of the molecules occur. It has been extensively studied in how far the
system can be influenced by such noise due to the feedback circle
\cite{arkinross,shea,ko,peccoud,mcadamsarkin,cook,bialek,sneppen}.
Note that this aspect of noise enters only in the number of involved molecules
as a function of time.

On the other hand, it might be argued that the spatial distribution of these
molecules, governed by Brownian motion, may give additional cause for the
influence of noise. Usually, it is replied that the system is well-stirred,
and therefore the molecules are always close enough to the switch on the DNA
such that the spatial effects can be neglected. With the typical diffusion
constant $K\sim 2\cdot 10^{-6}{\rm cm}^2/{\rm sec}$ for a molecule of 50{\AA}
size, the average diffusion time it takes to cross the cell is of the order of
one to a few msec. Processes affected by genetic switching occur on time scales
of the order of cell division of one or a number of cell generations, typically
tens of minutes or longer. The molecules, on this long time scale, are
therefore well-mixed within the cell. The effects of noise on the switching
process can, under this well-stirredness assumption, be considered directly
in the biochemical circuit
\cite{arkinross,shea,ko,peccoud,mcadamsarkin,cook,sneppen},
or it can be thought of as an activation process \cite{bialek}.

However, we will show that the involved system parameters are not consistent
with the above reasoning, and that well-stirredness alone is not sufficient
to obtain a complete picture of the process. In contrast, we give evidence that
the overall process consists of a large number of subprocesses during which
there is an ongoing competition between the protagonist and antagonist
molecules which in turn gives rise to the influence of spatial fluctuations on
genetic switching. Essentially, the reason for this claim is that the cell
volume is large in comparison to both the size of the messenger molecules
involved and the {\em van der Waals (vdW) interaction radius} around the
operator sites on the DNA. If one divides the cell volume into compartments
of the size of this vdW radius, the occupation of individual compartments by
the few molecules in the entire cell shows large fluctuations in time.

In what follows, we exclusively consider the effects of spatial fluctuations
in the above compartment picture. We distinguish the non-cooperative and the
cooperative cases, and our model switch is supposed to work according to the
following simple rules. (i) The non-cooperative switch gives rise to the state
of lysogeny if one R is bound to the operator. This R can dissociate from the
operator with some time constant, and it can be replaced by another molecule,
R or $\overline{{\rm R}}$, which is within the vdW interaction volume
(IV). If either another R substitutes the dissociated one, or the operator
remains vacant, the dormant state is preserved. The switch is flipped, and
lysis initiated, if eventually an $\overline{{\rm R}}$ molecule binds to
the operator site. (ii) In the cooperative case, two R's can bind to the
operator. In this configuration, the first facilitates the binding of the
second. Only if both R's dissociate from the operator and are eventually
replaced by one $\overline{{\rm R}}$, the switch flips towards lysis. Thus,
after full dissociation of the R (or R's) from the operator, the question
for both cases is
whether there is at least one R and/or $\overline{{\rm R}}$ within the
IV. If only one species is present, we assume that binding
of one molecule of this species necessarily occurs. If both species are
present, we introduce a $(1-\chi)$ factor in favour of R binding (and $\chi$
in favour of $\overline{{\rm R}}$ binding).

Consequently, our model can be rephrased as a {\em renewal process} in the
following sense. As the distribution of messenger molecules outside the IV
is irrelevant, the occupation of the IV can be regarded independent of the
previous occupation after the diffusion time it takes a molecule to cross
the IV, the {\em renewal time} $\delta t$. With the typical vdW
radius of 100{\AA}, we find with the above $K$ that $\delta t\sim 10^{-6}$
sec. Keeping track of the systems at ``stroboscopic'' times $\delta t,
2\delta t,\ldots$, we can employ a simple statistical analysis. The basic
ingredients are the probabilities $\Pi_0$ and $\Lambda_0$ that neither
R nor $\overline{{\rm R}}$ is inside the IV, and that there is at least
one such molecule present, $1-\Pi_0$ and $1-\Lambda_0$. If $p=\{\mbox{IV}
/\mbox{cell volume}\}$ is the probability that a single molecule is
within IV, and there are $N_{\rm R}$ and $N_{\overline{{\rm R}}}$
molecules of either species within the cell, we find $\Pi_0=(1-p)^{N_R}$
and $\Lambda_0=(1-p)^{N_{\overline{{\rm R}}}}$ \cite{remc}.

Let us collect some relevant numbers. The radius of {\em E.coli} is of the
order of 1$\mu$m, and the free volume in which the messenger molecules diffuse
within the cell is $\sim 1\mu$m$^3$ \cite{rema}. Comparing to the vdW radius,
we obtain $p\sim5\cdot 10^{-5}$. Note that for these numbers,
the probability that {\em none out of 100 molecules} is within the IV,
$(1-p)^{100}\approx 99.5\%$, is still very close to 1. The presence of such
small numbers which give rise to the fact that the associated probabilities
are either close to 0 or to 1 is the reason for the relevance of spatial
fluctuations.

Consider first the non-cooperative case. Assume that the bound R molecule
dissociates with the characteristic time scale $\tau$. Excluding that the
molecule does not immediately rebind, one witnesses a competition between
the two kinds of messenger molecules which can possibly bind to the relevant
operator sites. This competition is characterised through the four events
$P_1=\Pi_0\Lambda_0$, $P_2=(1-\Pi_0)\Lambda_0$, $P_3=(1-\Pi_0)(1-\Lambda_0)$,
$P_4=\Pi_0(1-\Lambda_0)$ which define the joint presence or absence of the
two types of molecules. These four configurations can be subdivided into
those which leave the genetic switch in the dormant mode, i.e., which prevent
an $\overline{{\rm R}}$ molecule from binding to the operator site, and those
which lead to $\overline{{\rm R}}$ binding to its operator. The former
comprise $P_1$ and $P_2$, the latter is given through $P_4$. In turn, $P_3$
defines a mixed state whose mean outcome will be $(1-\chi)$ in favour of R
binding, and $\chi$ in favour of $\overline{{\rm R}}$ binding. The
probability for inhibition is thus $P_{\rm inhib}=P_1+P_2+(1-\chi)P_3$,
the one for lysis is $P_{\rm lys}=1-P_{\rm inhib}$.

The whole process can
therefore be stripped down to the occurrence of a number $i$ of inhibition
events, terminated by a step leading to lysis, i.e., lysis will eventually
occur according to a sequence $P_{\rm inhib},P_{\rm inhib},\ldots,P_{\rm 
lys}$ with joint probability $P_{\rm inhib}^iP_{\rm lys}$ and
normalisation ${\cal N}=P_{\rm lys}/(1-P_{\rm inhib})=1$. The mean
time to obtain lysis according to this diffusion picture is
$\langle\delta t\rangle=\delta t\sum_{i=0}^{\infty}(i+1)P_{\rm
lys}P_{\rm inhib}^i$, resulting in $\langle\delta t\rangle=\delta t/P_{\rm
lys}$. The quantity $\langle\delta t\rangle$ increases with growing $N_{\rm
R}$, with decreasing $N_{\overline{{\rm R}}}$, or with decreasing $\chi$, as
it should. $\langle\delta t\rangle$ is the time due to the diffusion
renewal process. To obtain the overall characteristic time for lysis, we have
to add the binding times of order $\tau$. This delay can be included through
the average number of steps $\langle i\rangle=\sum_{i=0}^{\infty}iP_{\rm
lys}P_{\rm inhib}^i=P_{\rm inhib}/P_{\rm lys}$, weighted by the probability
$(P_2+(1-\chi)P_3)$ that a renewal step actually involves a rebinding of an R.
Multiplied by $\tau$ and added to $\langle\delta t\rangle$, this leads to
the characteristic lysis time
\begin{equation}
\label{Tnco}
T_{\rm nc}^{\rm lys}=\frac{\tau(P_2+(1-\chi)P_3)}{P_{\rm inhib}P_{\rm lys}}
+\frac{\delta t}{P_{\rm lys}}
\end{equation}
which will be discussed in comparison to the time scale in the cooperative case.

As mentioned, the cooperative scenario involves
two R molecules. If one is already bound to an operator site, it
facilitates the binding of another R molecule to the second operator site
reserved for R such that $\chi\approx 0$. Usually, two R's are bound. The
antagonist molecule $\overline{{\rm R}}$ can only bind and initiate the
divergence to the lytic track if both R sites are vacated.
I.e., if the one R dissociates {\em and} does not rebind during the
dissociation time of the second R. Moreover, one has to consider that not
each time both R's are dissociated, $\overline{{\rm R}}$ binds. In fact,
some R molecule can bind to the operator sites and restart the dissociation
process. As usually more R than $\overline{{\rm R}}$ are within the cell,
this case occurs more often, on average. Thus, if the characteristic time
$\tau^{II}$ for the dissociation of the second R is large in comparison to
the renewal time $\delta t$, a sufficiently high number of R molecules makes
it rather improbable that the R-related operator sites
remain unoccupied long enough as to allow for the complete dissociation of R
to occur: the characteristic time for lysis in the cooperative case
should be considerably higher than for the non-cooperative case \cite{REMM}.

To quantify this cooperative process, let us assume that $s=\tau^{II}/\delta
t$ is the number of renewal steps corresponding to the dissociation time of
the second R that is still bound. After dissociation, $1-\Pi_0$ defines the
probability that, in one given renewal step, an R
molecule binds to the vacant operator site and reconstitutes the initial
configuration with two R's bound to the DNA. The probability that during
$\delta t$ no such reconstitution occurs is given by $\Pi_0$.
The probability that reconstitution occurs in less than $s$ renewal
steps is then described by the combined process
$\overline{\eta}=(1-\Pi_0)\sum_{i=0}^{s-1}\Pi_0^i$,
obtaining $\overline{\eta}=1-\Pi_0^s$. The target
process for finding the possibility for lysis thus corresponds to one of the
following cascades of events, $\eta,\overline{\eta}\eta$, $\ldots$, $\overline{
\eta}^i\eta,\ldots$ where $\eta\equiv1-\overline{\eta}$. I.e., a certain number
of superprocesses $\overline
{\eta}$ occurs during which reconstitution takes place, and finally no R
replaces the dissociated first R until the second R dissociates, too.
The associated mean number of ``superprocesses''
$\overline{\eta}$ is $\langle i\rangle_{\overline{\eta}}=\overline{\eta}/(1-
\overline{\eta})$. In order to
estimate the characteristic time connected to this process, we have to include
two contributions. The first is the average time consumed by an $\overline{
\eta}$ superprocess; that is, $\tau_{\overline{\eta}}=
\delta t\sum_{i=0}^{s-1}(i+1)\Pi_0^i=$ $\delta t
(1-\Pi_0^s+s\Pi_0^{s+1}-s\Pi_0^s)/(1-\Pi_0)$.
The second is the dissociation time $\tau^I$ elapsing
after each re-binding of R. Finally, as 
not each complete dissociation of the two R molecules leads to a successful
binding of the antagonist $\overline{{\rm R}}$ molecule,
we obtain the characteristic time scale
\begin{equation}
\label{Tco}
T_{\rm c}^{\rm lys}=\frac{\langle\tau_{\rm diss}^{I\& II}\rangle
(P_2+(1-\chi)P_3)}{P_{\rm lys}P_{\rm inhib}}+\frac{\delta t}{P_{\rm lys}}
\end{equation}
for the occurrence of cooperative lysis. In Eq. (\ref{Tco}), the
time constant for complete dissociation of both R's
is given by
$\langle\tau_{\rm diss}^{I\& II}\rangle =\langle i\rangle_{\overline{
\eta}}\tau_{\overline{\eta}}+\left(\langle i\rangle_{\overline{\eta}}+1
\right)\tau^I.$ It is due to the
additional weighting through $\langle i\rangle_{\overline{\eta}}$ that
$T_{\rm c}^{\rm lys}$ exceeds $T_{\rm nc}^{\rm lys}$ considerably.

\begin{figure}
\unitlength=1cm
\begin{picture}(6,5.4)
\put(-4,-7.4){\includegraphics{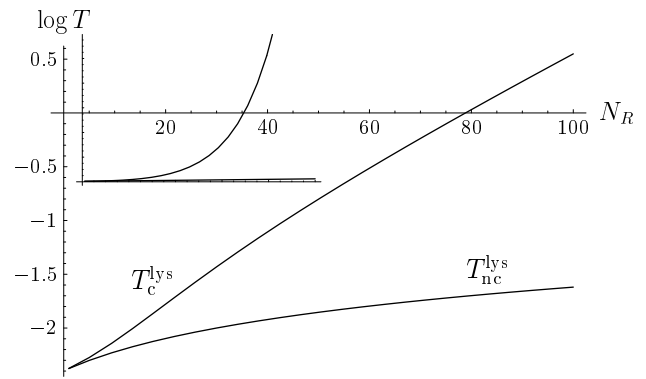}}
\end{picture}
\caption{Characteristic lysis times $T_{\rm nc}^{\rm lys}$ and $T_{\rm c}
^{\rm lys}$ as function of $N_{\rm R}$. $T_{\rm c}^{\rm lys}$ grows
exponentially with $N_R$, i.e., it corresponds to an almost linear increase
in the logarithmic case. Conversely, $T_{\rm nc}^{\rm lys}$ grows linearly.
The drastic difference between these two patterns is demonstrated in the 
inset with linear axes.
Parameters: $\delta t=10^{-6}$sec, binding/dissociation times $10^{-3}$sec,
$N_{\overline{{\rm R}}}=5$.
\label{fig1}}
\end{figure}

Both characteristic times can be evaluated numerically. Essentially,
the non-cooperative lysis time $T_{\rm nc}^{\rm lys}$, for a
fixed number $N_{\overline{{\rm R}}}$, grows almost linearly in $N_{\rm R}$,
compare Fig. \ref{fig1}. In contrast, $T_{\rm c}^{
\rm lys}$ grows almost {\em exponentially} for fixed $N_{\overline{{\rm R}}}$,
eventually reaching extremely large values for higher $N_{\rm R}$.
For $N_{\rm R}=1$, both characteristic times coincide,
as they should. Conversely, for fixed $N_{\rm R}$, both characteristic
lysis times fall off like a power law for increasing $N_{\overline{{\rm R}}}$.

These results are in qualitative agreement to those obtained from models
considering exclusively number fluctuation: cooperativity enhances the
accuracy of the system (the resistance against noise) exponentially,
compare, e.g., \cite{bialek}.
As our diffusion based model can lead to significantly large lysis
times which can be of the same order of magnitude as the results from
the number fluctuation models or even larger, depending on the assumed
parameters, it may not be sufficient to consider number fluctuations
only. In particular, it has been demonstrated that the well-stirredness
assumption is no sufficient a priori condition to exclude the spatial
inhomogeneities arising from the spatial diffusion of the molecules.
This is based on the fact that each dissociation-rebinding process
is influenced by the fairly high probability that no molecules are in the
interaction volume during a renewal step.

It should be emphasised that our results are sensitive to the very numbers
which are assumed for obtaining estimates for the characteristic lysis times.
A small variation of these numbers can lead to a large change in the final
result, so for a given system the parameters should be carefully verified
before estimates like the ones obtained herein are calculated.

The basic ingredient of our model is the separation of the entire free cell
volume into a bath constituted by the free diffusion volume, and into the
IV. Due to this assumption, the very configuration outside
the IV can be neglected. Consequently, for Monte Carlo
simulations of the combined process in which both the number of molecules
{\em and} their spatial variation are random,
the concept of the IV versus the free volume might prove
useful in stripping off the unnecessary details and lowering the computation
time.

Our renewal time scenario relies on the existence of a more
or less homogeneous distribution of the messenger molecules throughout
the free diffusion volume such that the net exchange with the IV
is approximately stationary. In prokaryotic cells, this assumption
should always be realised. It should also be valid in eukaryotes which 
feature a highly structured cell volume as long as there are no adsorption
processes at cellular membranes which lead to immobilisation of the
molecules according to a broad waiting time distribution which would give
rise to a diverging exchange rate \cite{report}.

By and large, in biophysics and biochemistry the role of noise in genetic
circuitry, and cellular systems as a whole, has been increasingly assessed.
This Letter shows that the spatial aspect of such fluctuations should not be
neglected a priori, and its relevance for the process should be checked.

We finally remark that genetic switches are paradigmatic systems at hand of
which effects of noise are studied. The developed renewal-diffusion scenario
therefore essentially pertains to
a large variety of systems and processes involving lowly populated species
which are spatially distributed and trigger followup processes on entering
some interaction zone, ranging from cellular feedback circles to clustering
of bacteria, or to animal populations. In conclusion, there is no a priori
well-stirredness condition for such systems.

RM thanks Peter Wolynes for helpful discussions at the
University of Illinois where part of this study was carried out. This work
was supported through the Deutsche Forschungsgemeinschaft (DFG) within the
Emmy Noether programme.

\begin{thebibliography}{99}

\bibitem{darwin} C. Darwin, {\em The origin of species} (John Murray,
London, 1859)

\bibitem{ptashne} M. Ptashne, {\em A Genetic Switch: Phage $\lambda$ and
Higher Organisms} (Cell Press/Blackwell, Cambridge, MA, 1992)

\bibitem{alberts} B. Alberts et al., {\em The molecular biology of the 
cell} (Garland, New York, 1994); S. R. Bolsover et al., {\em From genes
to cells} (Wiley, New York, 1997)

\bibitem{arkinross} A. Arkin, J. Ross and H. H. McAdams, Genetics {\bf 149},
1633 (1998); H. H. McAdams and A. Arkin, Ann. Rev. Biophys. Biomol. {\bf 27},
199 (1998)

\bibitem{vankampen} N. G. van Kampen {\it Stochastic Processes
in Physics and Chemistry\/} (North--Holland, Amsterdam, 1981)

\bibitem{shea} M. A. Shea and G. K. Ackers, J. Mol. Biol. {\bf 181}, 211
(1985)

\bibitem{ko} M. S. H. Ko, H. Nakauchi and N. Takahashi, EMBO J. {\bf 9},
2835 (1990); M. S. H. Ko, J. Theor. Biol. {\bf 153}, 181 (1991); BioEssays
{\bf 14}, 341 (1992)

\bibitem{peccoud} J. Peccoud and B. Ycart, Theor. Popul. Biol. {\bf 48},
222 (1995)

\bibitem{mcadamsarkin} H. H. McAdams and A. Arkin, Trends Genet. {\bf 15},
65 (1999); Proc. Nat. Acad. Sci. USA {\bf 94}, 814 (1997)

\bibitem{cook} D. L. Cook, L. N. Gerber and S. J. Tapscott, Proc. Nat.
Acad. Sci. USA {\bf 95}, 6750 (1998)

\bibitem{sneppen} E. Aurell, S. Brown, J. Johanson and K. Sneppen,
cond-mat/0010286

\bibitem{bialek} W. Bialek, cond-mat/0005235

\bibitem{remc} Here we assume that both kinds of molecules are of
comparable size. This is reasonable as both have to bind to similar
operator sites. It is, however, straightforward to include differences
in the size.

\bibitem{rema} Taking into consideration that a portion of the entire
cell volume is occupied by other functional units like vesicles etc.

\bibitem{REMM} Lysis can be prompted, for instance, by exposure of the
bacterium cell to UV light which leads to cleavage ot R molecules such
that they cannot bind to the DNA any longer. Thus, $\overline{\mbox{R}}$ can
substitute without competition with R, and induce lysis immediately.

\bibitem{report} R. Metzler and J. Klafter, Phys. Rep. {\bf 339}, 1
(2000)

\end{thebibliography}
\end{document}